\documentclass[pdflatex,sn-mathphys-num]{sn-jnl}

\usepackage[T1]{fontenc}
\usepackage{lmodern}
\usepackage[utf8]{inputenc}
\usepackage{graphicx}%
\usepackage{multirow}%
\usepackage{amsmath,amssymb,amsfonts}%
\usepackage{amsthm}%
\usepackage{mathrsfs}%
\usepackage[title]{appendix}%
\usepackage{xcolor}%
\usepackage{textcomp}%
\usepackage{manyfoot}%
\usepackage{booktabs}%
\usepackage{algorithm}%
\usepackage{algorithmicx}%
\usepackage{algpseudocode}%
\usepackage{listings}%
\usepackage{makecell}
\usepackage{float}
\usepackage{placeins}
\usepackage{comment}
\UseRawInputEncoding
\usepackage{natbib}

\usepackage[colorinlistoftodos]{todonotes} 

\theoremstyle{plain}%
\begin{document}

\title[Article Title]{Numerically optimized FROG results for the study of red-shifted spectra in multi-frequency Raman generation}


\author[1]{\fnm{Sakthi Priya} \sur{Amirtharaj}} 

\author[2]{\fnm{Zujun} \sur{Xu}} 

\author[2]{\fnm{Donna} \sur{Strickland}} 

\author[3]{\fnm{Borun} \sur{Chowdhury}} 

\author[4]{\fnm{Sagnik} \sur{Acharya}} 

\author[5]{\fnm{Priyam} \sur{Samantray}} 

\author[1]{\fnm{Anil} \sur{Prabhakar}} 

\author[5]{\fnm{Kisor} \spfx{Kumar} \sur{Sahu}} 

\author[6]{\fnm{Franz} \sur{Bamer}} 

\author*[5,6]{\fnm{S.} \sur{Swayamjyoti}} \email{swayamjyoti@iam.rwth-aachen.de} \email{swayam.sj@iitbbs.ac.in}

\affil[1]{\orgdiv{Department of Electrical Engineering}, \orgname{IIT Madras}, \orgaddress{\city{Chennai}, \postcode{600036}, \country{India}}}

\affil[2]{\orgdiv{Department of Physics and Astronomy}, \orgname{University of Waterloo}, \orgaddress{\city{Waterloo}, \postcode{382355}, \country{Canada}}}

\affil[3]{\orgdiv{} \orgname{Meta}, \orgaddress{\city{London}, \postcode{W1T 1FB}, \country{UK}}}

\affil[4]{\orgdiv{Department of Mechanical Engineering}, \orgname{Indian Institute of Science}, \orgaddress{\city{Bangalore}, \postcode{560012}, \country{India}}}

\affil[5]{\orgdiv{School of Minerals, Metallurgical and Materials Engineering}, \orgname{IIT Bhubaneswar}, \orgaddress{\city{Argul}, \postcode{752050}, \state{Odisha}, \country{India}}}

\affil[6]{\orgdiv{Institute of General Mechanics}, \orgname{RWTH Aachen University}, \orgaddress{\city{Aachen}, \postcode{52062}, \country{Germany}}}


\abstract{
   \hspace{15pt} 
   When multifrequency Raman scattering is driven in the transient regime by two chirped pump pulses, the resulting anti-Stokes orders exhibit asymmetric spectral broadening toward lower frequencies, leading to a characteristic double-peaked structure in each order. In this Letter, frequency-resolved optical gating (FROG) is used to investigate the spectral evolution of the first anti-Stokes Raman component.
   To interpret the observed features, we introduce a double-pulse interference model and employ an adaptive learning-based reconstruction algorithm using the Adam optimizer to retrieve the temporal field evolution. The simulation results show good agreement with the experimental measurements. Our analysis indicates that the observed red-shifted spectral component originates from linear Raman processes within the two-photon dressed-state framework.
}


\keywords{Frequency Resolved Optical Gating (FROG),  Multi-frequency Raman Generation, Machine Learning, Numerical Optimization}



\maketitle

\newpage

\section{Introduction}\label{sec1}

Laser pulses are of prime importance owing to their role in a wide array of applications, ranging from military to medical operations. Laser pulses with high peak power can reach durations on the order of femtoseconds and are referred to as ultrashort laser pulses. Such high-intensity fields often drive nonlinear interactions in optical media. A key feature of ultrashort laser pulses is their temporal and spectral coherence: since the pulse is composed of a broad frequency bandwidth, maintaining a well-defined phase relationship across this spectrum is essential. This coherence allows the pulse to remain localized in time and enables precise control over light-matter interactions.
The development of ultrashort laser pulses has been a significant breakthrough, enabling a broad spectrum of applications such as fast imaging \cite{silberberg_quantum_2009, sheetz_ultrafast_2009}, attosecond electron dynamics \cite{hu_attosecond_2006, kling_attosecond_2008, lepine_attosecond_2014}, and laser spectroscopy \cite{zewail_femtochemistry_2000, demtroder_laser_1996}. In addition, ultrashort pulses have found important use in laser eye surgery \cite{bille2011, chai_vivo_2010, plamann_ultrashort_2010, palanker_femtosecond_2010, krueger_first_2005, schumacher_femtosecond_2009, juhasz_corneal_1999}, surface texturing \cite{toyserkani_ultrashort_2015}, and material processing \cite{malinauskas_ultrafast_2016}. 

However, the duration of ultrashort pulses is often too short to be directly captured by sensors with limited temporal response. Consequently, indirect methods using computational algorithms are required for their retrieval. A further challenge arises from the fact that infinitely many pulse shapes can correspond to a single measured trace. To address these difficulties, the technique of Frequency-Resolved Optical Gating (FROG) \cite{trebino_measuring_1997, trebino2012, bendory_on_2017} is commonly employed. In FROG, autocorrelations between a pulse and its time-delayed replica are used to compute the intensity, frequency, and phase of the measured trace. The reconstruction is typically achieved using the Principal Component Generalized Projection Algorithm (PCGPA) \cite{kane_principal_2008}. An alternative approach, ptychographic FROG \cite{sidorenko_ptychographic_2016}, has also been developed, offering improved handling of Fourier transforms within the retrieval algorithm and enabling more robust pulse characterization.

Recently, a study \cite{zahavy_deep_2018} was carried out to reconstruct ultrashort pulses using deep learning. Initially, supervised learning was applied to computer-simulated data, yielding better performance than both ptychographic FROG and PCGPA FROG. However, this approach required training of 60000 pulses, which is a shortcoming from the experimental point of view. To mitigate this, an unsupervised procedure that does learning on the experimental traces was used alongside the supervised learning. Upon comparison with the reference trace, the reconstructed FROG trace by this method was found to generate the lowest error as compared to the trace obtained by PGCPA and ptychographic FROG methods. Several other studies \cite{kleinert_rapid_2019, krumbugel_direct_1996} have also explored the use of deep neural networks to recover the temporal shape of femtosecond pulses from dispersion-scan traces, as well as computational neural networks to reconstruct pulse intensity and phase directly. In a subsequent study \cite{ziv_deep_2020}, a novel technique was introduced to recover the amplitude and phase of ultrashort pulses in a single shot using a deep learning framework, thereby eliminating the need for iterative algorithms. In this approach, pulse reconstruction is achieved by training a neural network to invert the nonlinear interference pattern for pulse mapping. Notably, the method demonstrates robust performance even under conditions of low signal-to-noise ratio.

FROG investigations are highly effective for probing the dynamics of ultrafast pulses. Recently, Xu et al.\cite{xu_time-dependent_2023} employed cross-FROG analysis to address the long-standing question regarding the origin of the additional red-shifted peak observed when two linearly chirped pulses are coupled during transient Multi-frequency Raman Generation (MRG) experiments. MRG is a cascaded Raman scattering process that produces a broad spectrum of discrete Raman orders \cite{imasaka_generation_1989}. It is a promising technique for creating ultrashort pulses with high energy. While investigating MRG in the transient regime, Xu et al. observed that individual Raman orders display characteristic double-peaks as the instantaneous frequency separation of the two pump pulses was tuned to the red side of the Raman transition. The FROG study of the anti-Stokes orders revealed that linear Raman scattering from two-photon dressed states explains the red-shifting phenomenon. Although the theoretical framework was well established, reproducing the effect through simulations proved challenging and required the use of a rigorous FROG retrieval algorithm \cite{trebino2012}.

In this work, we extend the cross-FROG analysis using a numerical optimization approach. A series of experiments was conducted to investigate the factors contributing to the red-shifted spectrum, including the instantaneous frequency separation between chirped pulses, intensity-dependent Rabi frequency, pulse energy, and pulse intensity. We developed a double-pulse model incorporating the intensity-dependent Rabi frequency to describe the red-shifted spectral phenomenon. Building on these hypotheses, we introduce a numerical framework capable of reconstructing the experimental traces, thereby providing deeper insight into the mechanisms underlying the red-shifted Raman orders. Overall, this work demonstrates a pathway to improving FROG reconstruction for complex pulse structures when the underlying pulse type is known.

We started with 65 sets of experimental traces for the optimization problem. Consequently, the numerical model, i.e., deep learning/optimization methods, must be able to provide good results with limited data. At the same time, it should be computationally efficient, i.e., be able to converge faster with fewer iterations. Stochastic Gradient Descent (SGD) \cite{robbins_stochastic_1951} is well suited to these requirements, as it randomly samples data points from the full dataset in each iteration, reducing computational cost. However, instead of using the stochastic gradient descent, we have adopted a variant of SGD named ADAM \cite{kingma_2015} for our optimization routine, like in the studies mentioned earlier. ADAM maintains an adaptive learning rate based on the weights of each network. It adaptively updates the step size during the optimization instead of adjusting the parameters at the onset. It is computationally efficient, easy to implement, requires less memory, and fewer iterations for convergence. All these serve as motivation to adopt it for our current study. Also, in several studies \cite{jais_adam_2019}, the Adam optimization algorithm has been found to perform better.

The main objective of the present study is to seek more insight into the phenomenon of red-shifted spectrum observed during the transient MRG experiment \cite{xu_temporal_2018}. Firstly, an experimental setup is prepared to create two pulses of 786 nm and 837 nm
wavelength followed with experiments of varying instantaneous frequency separation and pump energies \cite{xu_time-dependent_2023}. We simulated the traces observed in the experiments to achieve a better understanding of the red-shifted spectrum. With the experimental trace as an input, ADAM is used to optimize the training parameters and develop the simulated trace. As described earlier, we adopted a double pulse model hypothesis for our numerical model. It is used in combination with the FROG algorithm to create a simulated trace. Further, the simulated trace is checked for accuracy with the original/experimental trace by a loss function (a root mean square error formulation function) within the ADAM optimization algorithm. At each step of ADAM optimization, the backward gradients, the updation of the training parameters, and the extraction of the simulation parameters corresponding to the minimal value of the loss function are computed. Note contrary to standard supervised learning paradigm where given a set of input output tuples $\{(x,y)\}$ one needs to find $\theta$ parameterizing a function $y=f_\theta(x)$, our method is more akin to "Deepfool" \cite{moosavi-dezfooli_deepfool_2015}, where given a set of outputs $\{y\}$ and a function $f(X)$, we want to find the corresponding inputs $\{x\}$, such that $y=f(x)$ for all the outputs in the training set. The optimization algorithm is ceased when several indicators pertaining to stability and errors of the simulated trace attain acceptable values. 

The paper is organized as follows. In Section \ref{sec2}, a brief introduction is provided to the experimental setup for the MRG in detail. The details pertaining to the simulation methodology are explained in Section \ref{sec3}. It includes the ADAM optimization algorithm, the hypothesis, a flowchart highlighting their implementation, and an explanation of the same. In Section \ref{sec4}, we present the simulation results obtained by minimizing the mean-squared error (MSE) between the simulated and experimental traces.
Finally, section \ref{sec5} reports the conclusions and perspectives of the simulation of our study. In section \ref{sec:ack} and section \ref{sec:appendix}, the acknowledgments and the appendix are discussed.

\section{Experimental Setup}\label{sec2}

Figure 2 shows a schematic of our experiment setup, based on our previous study \cite{xu_time-dependent_2023}. A dual-wavelength Ti: sapphire laser produces a pump beam at 786~nm and a Stokes beam at 837~nm for the excitation of MRG \cite{zhang_2000}. The two beams are passed through a grating compressor, which stretches the pulses independently. The compressor also includes a tunable mirror that is used to adjust the time delay between the two beams, which in turn modifies the instantaneous frequency separation between the pulses. Sulfur hexafluoride (SF\textsubscript{6}) in the hollow-core fiber is used as the Raman medium, with the frequency separation between the two beams matching the Raman frequency of SF\textsubscript{6} (23.25 THz). The hollow-core fiber is 500~cm long with an inner diameter of 150~$\mu$m. The pump and Stokes energies reaching the Raman medium are adjusted using a half-wave plate and a polarizer. The Stokes beam and the MRG output are then fed into a cross-FROG setup for analysis. The output from the hollow-core fiber is collimated and sent to a cross-FROG setup for pulse characterization, with the Stokes beam used as the reference pulse. This enables simultaneous investigation of multiple anti-Stokes orders. In this work, we focus on the first anti-Stokes order.

\begin{figure}[ht]
    \centering
    \includegraphics[width=0.6\textwidth]{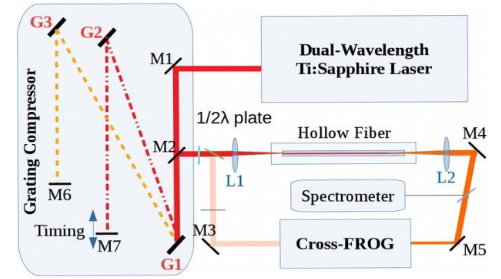}
    \caption{A schematic depiction of the experimental protocol. G, L and M stand for Grating Lens and Mirror, respectively. Figure adapted from \cite{xu_time-dependent_2023}.}
\end{figure}

\section{Modelling Methodology}\label{sec3}
\subsection{Numerical models}
\subsection*{Double pulse model}
For the numerical model, we need a theory that best describes the MRG in the transient regime. For steady-state pumps, Hickman's theory is widely used to describe the MRG process. This approach treats MRG within a multi-wave coupling picture, in which all Raman sidebands interact simultaneously~\cite{hickman_theory_1986}. The frequency of the $j$-th Raman component is expressed as $\omega_j = \omega_0 + j\omega_R$, where $\omega_0$ denotes the pump frequency and $\omega_R$ corresponds to the Raman shift. Within the two-photon Bloch formalism~\cite{rickes_efficient_2000}, the resonant Raman interaction is characterized by a generalized Rabi frequency $\Omega'$, defined as
\begin{equation}
\Omega'^2 = \Omega^2 + \Delta^2,
\label{eq:omega1}
\end{equation}
where $\Omega$ represents the on-resonance Rabi frequency and $\Delta$ denotes the effective detuning. These quantities are given by

\begin{equation}
\Omega e^{i\theta} = \frac{\alpha_{12}}{2\hbar} \sum_j V_j V_{j-1}^*,
\label{eq:omega}
\end{equation}

\begin{equation}
\Delta = \frac{\partial \theta}{\partial t} + \frac{2\pi (\alpha_{22} - \alpha_{11}) I}{\hbar c} + \delta \omega,
\label{eq:delta}
\end{equation}

\begin{equation}
I = \frac{c}{8\pi} \sum_j V_j V_j^*,
\end{equation}
where $V_j$ denotes the complex amplitude of the $j$-th Raman order,  $\alpha_{ij}$ are the transition moments, $I$ is the total intensity of all orders, and $\delta\omega$ is the detuning between the frequency separation of the pumps and the Raman frequency. 

Equations~\ref{eq:omega1}-\ref{eq:delta} indicate that the generalized Rabi frequency $\Omega'$ depends implicitly on the pulse energy through both $\Omega$ and $\Delta$. Increasing the pulse intensity enhances the coupling strength and modifies the effective detuning, thereby increasing $\Omega'$. Since the present work focuses on transient MRG driven by ultrashort pulses, the temporal envelope of the excitation must be explicitly included. Consequently, the generalized Rabi frequency becomes time dependent:


\begin{equation}
\Omega'^2(t) = |\Omega(t)|^2 + \Delta^2(t) 
\label{eq: omega'}
\end{equation}

In principle, an exact evaluation of $\Omega'(t)$ requires inclusion of all Stokes, anti-Stokes, and pump fields. However, such a treatment leads to a rapidly increasing computational burden as the number of coupled frequency components grows. Given that the experiment employs two dominant pulses corresponding to the pump and Stokes, the model is simplified by retaining only these contributions. As evident from Eq.~(\ref{eq:delta}), the detuning $\Delta$ comprises three distinct terms: the intrinsic Raman detuning $\delta\omega$, the temporal derivative of the Rabi phase $\theta$, and the two-photon Stark shift arising from the difference in polarizabilities $\alpha_{22}$ and $\alpha_{11}$ weighted by the total intensity. For the Raman medium used in this study (SF$_6$), previously reported polarizability values indicate that the contribution of the Rabi frequency dominates over the Stark shift term~\cite{xu_time-dependent_2023}, and thus primarily determines the magnitude of $\Omega'$.


To model the experimentally observed first anti-Stokes signal, a double-pulse representation is employed in conjunction with ADAM and FROG algorithms to generate simulated spectrograms. The anti-Stokes field is described as the coherent superposition of two Gaussian pulses, $E_1$ and $E_2$, with distinct center frequencies:
\begin{equation}
E \left(t\right) = E_{1} \left( T_1, f_1, A, \phi_1, \tau \right) + E_{2} \left( T_2, f_2, B, \phi_2, \tau \right)
\end{equation}
Here, $T_i$ denotes the pulse duration, $f_i$ the central frequency, $A$ and $B$ the field amplitudes, $\phi_i$ the carrier phase, and $\tau_i$ the temporal delay of each pulse ($i = 1,2$). Each Gaussian pulse is expressed as
\begin{equation}
E = A \cdot \exp\left(-2 \ln\left(2\right)  \left(\frac{t - \tau}{T}\right)^2\right) \cdot \exp\left(j2\pi f \left(t - \tau\right) - j\phi\right)  
\end{equation}
where the factor $2\ln 2$ ensures that the parameter $T$ corresponds directly to the full width at half maximum (FWHM) of the pulse envelope. This convention allows $T$ to represent a physically meaningful pulse duration rather than an arbitrary width parameter, thereby improving interpretability and consistency within the model.

Two simulation models were adapted - one with an ideal linear shift for the Raman pulse, and another using the experimentally collected pump data. 

\subsubsection*{Intensity-dependent phase of the red-shifted pulse (Beta-I)}
Intensity-dependent phase is incorporated into the double pulse mode, and it is referred to hereon as the beta-I model. In this model, the Raman phase follows a quadratic dependence on time, such that the instantaneous frequency would be linear with time. The red-shifted phase is a sum of the Raman phase, the time integral of the time-dependent generalized Rabi frequency shift, and a constant phase difference. The phases in the MRG model are defined as:
\begin{align}
\phi_1 &= a_2 \cdot \left(t-\tau\right)^2 \,, \\
\phi_2 &= \phi_1 + \int \beta \cdot I \, dt + \Phi 
\end{align}

These phases are used to compute the energy fields $E_1$ and $E_2$:
\begin{align}
E_1 &= E_{\text{Gaussian}}\left(T_1, f_1, A, \phi_{1}, \tau \right) \\
E_2 &= E_{\text{Gaussian}}\left(T_2, f_2, B, \phi_{2} , \tau\right)
\end{align}
where  $\Phi$ is a variable phase term, to account for phase shifts and chirp effects in the experimental setup.

\subsubsection*{Beta-I model with pump data}
\label{sec:pump}
In this model, the Raman phase takes the phase of the pump, which is experimentally measured for more realistic calculations. The red-shifted phase is defined as in the beta-I model. 
\begin{align}
\phi_1 &= \phi_{pump}  \\
\phi_2 &= \phi_1 + \int \beta \cdot I \, dt + \Phi
\end{align}

These phases are used to compute the energy fields $E_1$ and $E_2$:
\begin{align}
E_1 &= E_{\text{Gaussian}}\left(T_1, f_1, A, \phi_{1}, \tau \right) \\
E_2 &= E_{\text{Gaussian}}\left(T_2, f_2, B, \phi_{2}, \tau\right)
\end{align}
where  $\Phi$ is a variable phase term.

\subsection{Numerical optimization}
In this work, we obtained experimental traces, represented as intensity field plots during multi-frequency Raman scattering experiments. The primary goal of our modeling is to reconstruct (or simulate) these experimental traces through optimization, in order to estimate the values of certain parameters that minimize the deviation, referred to as delta. These parameters include the frequencies (f0, f1), time durations (T1, T2), amplitudes (A, B), time delays, and phase factors (a2, a3) of the pump and Stokes pulses. We refer to these as the simulation parameters.
Here, delta serves as a measure of the difference between the simulated and experimental traces. In this section, we briefly describe our optimization technique, its formulation, and the complete modeling methodology, along with flowcharts (see Section \ref{sec6}) illustrating its implementation in our developed codebase.

\begin{figure}[ht]
    \centering
    \includegraphics[width=1.0\textwidth]{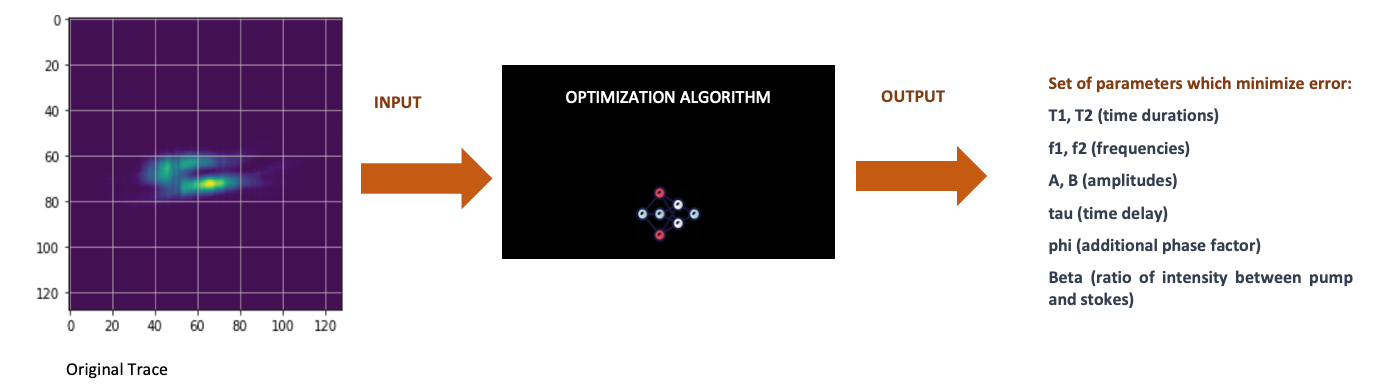}
    \caption{Overview of our approach.}
\end{figure}

\subsubsection*{Choice of ADAM} 
In this work, we employed the ADAM optimization algorithm \cite{kingma_2015}, an enhanced stochastic gradient-based method introduced earlier in the manuscript. ADAM was selected because it integrates adaptive learning-rate control with momentum-like updates for each network parameter. Specifically, it maintains exponentially decaying estimates of both the first and second moments of the gradients. The latter plays a role analogous to the mechanism used in RMSProp \cite{tieleman_2012}, where gradient scaling is achieved through a running average of squared gradients, while the former resembles momentum-based methods \cite{Ning_1999} that accumulate gradient history to stabilize and accelerate convergence.

By combining these two strategies within a single framework, ADAM dynamically adjusts the effective step size throughout the optimization process rather than relying on a fixed learning rate.
It is computationally efficient, easy to implement, requires less memory, and fewer iterations for convergence. All these serve as motivation to adopt it for our current study. Moreover, in several studies (\cite{jais_2019} and \cite{khan_2020}), the Adam optimization algorithm has been found to perform better. 

\subsubsection{Formulation} 
In our current formulation, we organize the numerical workflow into three key Python modules, namely froggen.py, trainer.py, and diffarray.py, in addition to the primary script AllRuns.py. The details of these Python codes and their implementation are explained in the section \ref{sec:appendix}. 

We begin with an initial guess for the simulation parameters. A reasonable range for each parameter is selected based on previous results from the FROG retrieval algorithm \cite{xu_red-shifted_2021} and expected values from the experiment. The values for the pulse durations and the additional phase term $\phi$ are strictly constrained within this range throughout the ADAM optimization routine. The remaining parameters are bounded only during the initial guess. We have observed that the outcome of the optimization is influenced by the initial parameter ranges, making it essential to choose ranges that are physically reasonable. Because the phases $\phi_i$ are periodic quantities, we enforce circular consistency during optimization. Each update to $\phi_i$ is wrapped modulo $2\pi$ to ensure
that the parameter remains on its natural circular domain and does not encounter discontinuities at the boundary.

The script froggen.py generates the FROG trace from the inputted values of the simulation parameters through the FROG algorithm, which involves the Fast Fourier Transform to generate a matrix consisting of the real and imaginary parts of the simulated trace. The Beta-I model computes the energy fields and subsequently the output matrix, leading to the simulated trace. For more details, please refer to section \ref{sec3} and sub-section \ref{subsec2}.

The trainer.py executes the ADAM optimization algorithm iteratively to compute the loss $\mathcal{L}$, i.e. difference between the matrix representing the simulated and experimental trace. 
\begin{align}
\mathcal{L}
= \frac{1}{N} \sum_{i,j} \left( F_{\text{sim}}(i,j) - F_{\exp}(i,j) \right)^2 ,
\end{align}

Also, a normalized metric is defined for the loss, termed as "Diff" value $\delta$. Diff value is given by,
\begin{align}
\tilde{F}_{\exp}(i,j) = \frac{F_{\exp}(i,j)}{\max_{i,j} F_{\exp}(i,j)}, \\
\qquad
\tilde{F}_{\text{sim}}(i,j) = \frac{F_{\text{sim}}(i,j)}{\max_{i,j} F_{\text{sim}}(i,j)}.\\
\delta 
= \frac{\displaystyle \sum_{i,j} \left( \tilde{F}_{\exp}(i,j) - \tilde{F}_{\text{sim}}(i,j) \right)^2}
{\displaystyle \sum_{i,j} \tilde{F}_{\exp}(i,j)^2}.
\end{align}

Also, the matrix containing the simulated trace generated by the froggen.py is further used in diffarray.py to compute the delta between the simulated and experimental trace. For more details on trainer.py, please refer to sub-section~\ref{subsec1}.

During each iteration of the optimization, the backward gradients, updation of the simulation parameters, and extraction of the simulation parameters corresponding to the minimal value of the loss parameter are computed just like any regular optimization routine. The optimization algorithm is ceased when several indicators pertaining to stability and errors of the simulated trace attain acceptable values.

In order to gain insight into the entire modeling methodology, please see Section \ref{sec3}.

\section{Results and discussion}\label{sec4}

In the initial series of measurements, we investigated how variations in pump pulse energy influence the observed red-shifting phenomenon. To begin with, the temporal and spectral characteristics of the pump pulses were characterized using an auto-FROG configuration. These measurements serve as the reference fields for subsequent cross-FROG characterization of the first anti-Stokes Raman signal.  The retrieved parameters of the pump pulses are summarized in Table~\ref{table: pump pulse energy}. The two pump beams are centered at wavelengths of 786 nm and 836 nm, respectively.

\begin{table}[!ht]
\caption{Pumps for the energy scan MRG experiments}
\begin{tabular}{llll}
& Center-wavelength & Pulse duration & Chirp rate (linear approximation) \\
Pulse & 786.2 nm & 902 fs & 1.60 THz/ps \\
Stokes & 836.5 nm & 831 fs & 1.50 THz/ps \\
\end{tabular}

\label{table: pump pulse energy}
\end{table}

Following this characterization, cross-FROG spectrograms of the anti-Stokes Raman signal were acquired, with the Stokes pulse serving as the reference, over a range of pump energies. The first anti-Stokes Raman order was selected for detailed analysis in all subsequent measurements.  The relative temporal delay between the two pump pulses was fixed at the value that maximized the number of generated orders when operating at the maximum available total pump energy of 2.2 mJ. Throughout the energy scan, the pumps were maintained at a frequency separation of 23.25 THz and were prepared with comparable linear chirp rates. Owing to the upper limit of 2.2 mJ on the total pump energy, measurements were performed in discrete steps of 0.5 mJ.

\begin{figure}[ht]
    \centering
    \includegraphics[width=1.0\textwidth]{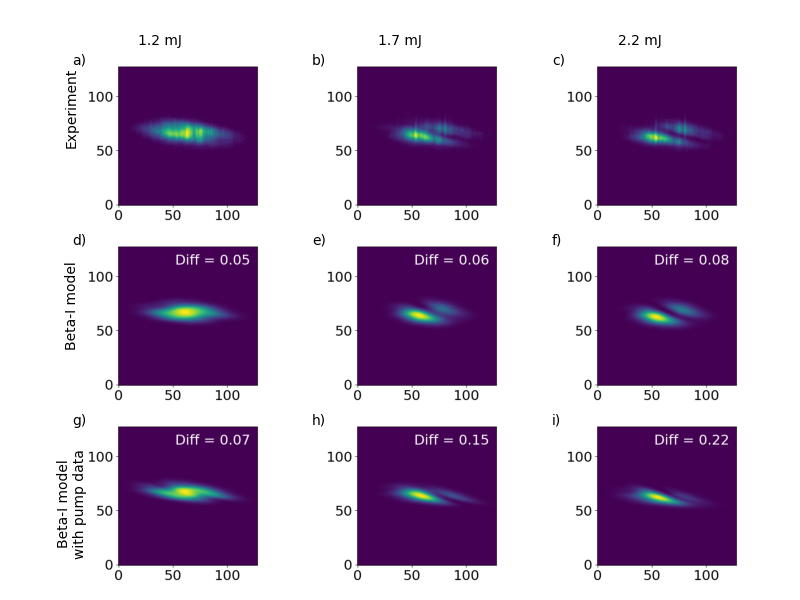}
    \caption {Cross-FROG traces comparison. Experimental traces (top row); Reconstructed FROG traces by beta-I model (middle row); Reconstructed FROG traces by beta-I model with pump data (bottom row). The columns represent the energy scans with different pump energies -  1.2~mJ, 1.7~mJ, and 2.2~mJ from left to right. Diff value represents the FROG reconstruction error. A small diff indicates better matching. }
    \label{fig: frog}
\end{figure}

From the experimental FROG trace, we use the ADAM optimization approach to arrive at the electric field parameters that minimize the FROG reconstruction error. The reproduced FROG traces along with the experimental FROG trace using both the beta-I model and the beta-I model with pump data are shown in Fig.\ref{fig: frog}. The FROG reconstruction error, or the mean-squared error, is represented by the diff value. A diff value less than 0.08 is assumed to be a good fit based on statistical results from the FROG retrieval algorithm \cite{xu_time-dependent_2023}. We found that the beta-I model is performing well in reproducing the experimental trace as compared to the beta-I model with pump data. This can be attributed to the irregularities in the pump data. The pump phase is quite noisy at the beginning and end of the time trace; consequently, the instantaneous plot captures these irregularities. This gives rise to increased error compared to the beta-I model. 

We had 65 experimental data sets, including the energy scans and the time delay scans (which will be discussed later in this section). Beta-I model has performed well with 71\% of the dataset with error values less than 0.08. This is achievable with a one-time run of our code with GPU implementation and is reproducible. This is far efficient compared to the earlier approach of the FROG retrieval algorithm, which is heavily dependent on the initial conditions, where a user has to manually play with a range of initial conditions before arriving at the optimal solution. BetaI model with pump data has 29\% of the dataset with error values less than 0.08. But, this can be attributed to the noise in the pump data as explained previously.

\begin{figure}[ht]
 \centering
  \includegraphics[width=1.0\textwidth]{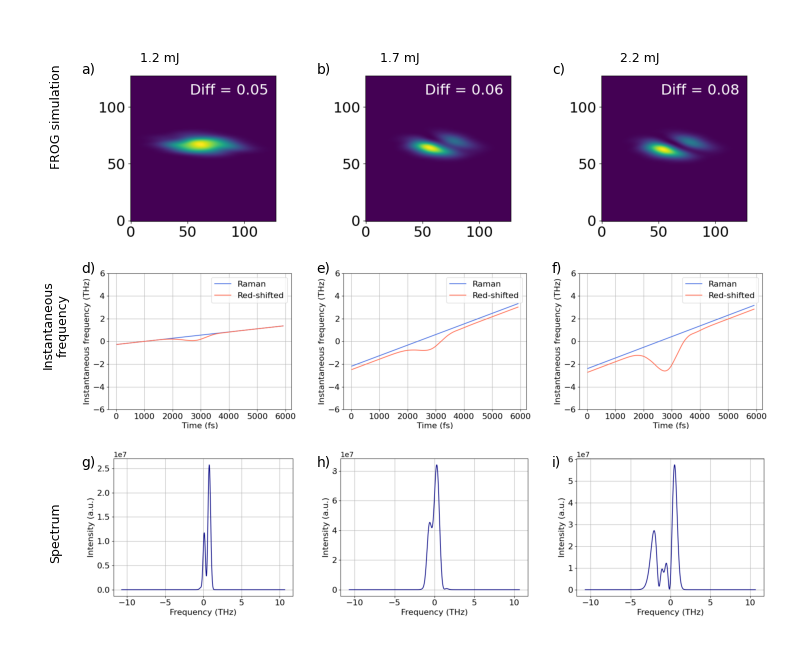}
 \caption{Simulated FROG traces (top row) for the first anti-Stokes Raman component at three different total pump energies: 1.2~mJ, 1.7~mJ, and 2.2~mJ, shown from left to right. For each energy, the corresponding instantaneous frequency of the Raman signal, including the red-shifted contribution (middle row), and the associated spectral profile (bottom row) are also presented.}
 \label{fig: energy scan frog}
 \end{figure}

The FROG reconstruction provides us with the parameters of the electric field of the combined double-pulse model with Raman and red-shifted peak. It can be further used to understand the behavior of the MRG orders. We now discuss the detailed results from the energy scans (Fig.~\ref{fig: energy scan frog}). From the individual Raman and red-shifted pulses, we can compute the instantaneous frequency over time, which is the derivative of the individual phases with time. The Raman pulse shows a linear trend in the instantaneous frequency vs time, as expected for a linearly chirped pulse. The red-shifted pulse, however, shows a dip in the instantaneous frequency response at half the pulse duration. This is on account of the Rabi-frequency shift as explained in the previous work of Xu et al. \cite{xu_time-dependent_2023}. As evident from the FROG trace, with increasing pump energy, the separation of the two peaks in the first anti-Stokes order is more clearly visible as in Fig.~\ref{fig: energy scan frog}i. The trend in the instantaneous frequency shows that the dip near half-time duration increases with the increase in pump energy. The overall pulse is a lot more complicated, but we could retrieve the features of the electric field from the instantaneous frequency calculations. As evident from the distinction of the two peaks and their phases, the first anti-Stokes Raman order with double peaks is a mixing of two pulses.


Next, we learn the effect of varying the instantaneous frequency separations. Towards that end, the pump pulses were characterized with auto-FROG measurement as before. They have similar pulse duration around 800 fs, and similar linear chirp rate around 1.2 THz/ps.

\begin{table}[ht]
\caption{Measurement of the two pumps (pump and Stokes) with auto-FROG for time delay scans.}
\begin{tabular}{llll}
& Center-wavelength & Pulse duration & Chirp rate (linear approximation) \\
Pulse & 786 nm & 781 fs & 1.18 THz/ps \\
Stokes & 836 nm & 830 fs & 1.23 THz/ps \\
\end{tabular}

\end{table}

\begin{figure}[ht]
 \centering
  \includegraphics[width=1.0\textwidth]{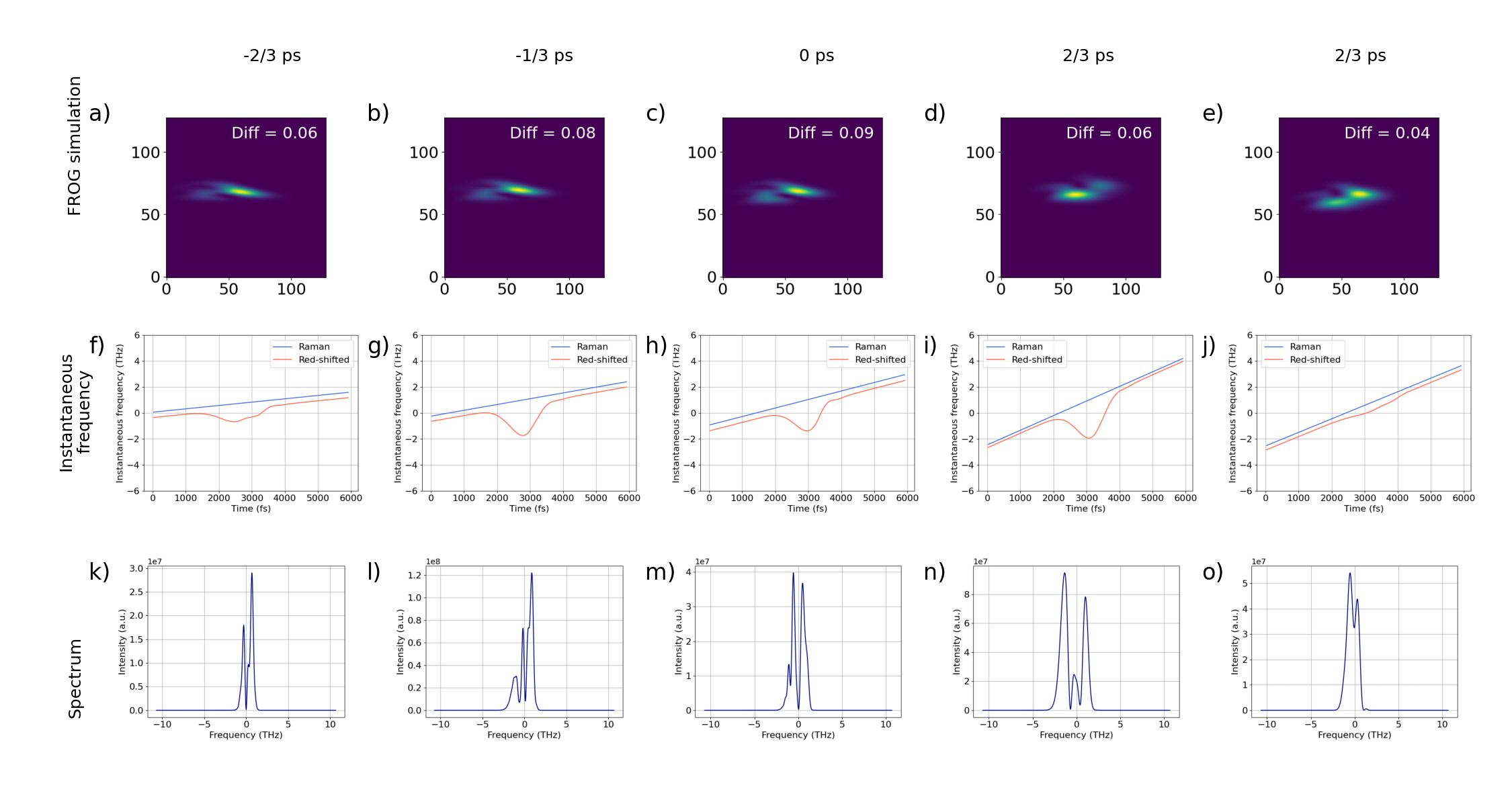}
  \caption{Simulated FROG traces (top row) for the first anti-Stokes Raman component obtained at different relative delays between the two pump pulses: $-2/3$~ps, $-1/3$~ps, $0$~ps, $1/2$~ps, and $2/3$~ps, displayed from left to right. For each delay setting, the corresponding instantaneous frequency evolution of the Raman and red-shifted pulse (middle row), along with the associated spectral profile (bottom row), is presented.}
 \label{fig: timedelay scan}
 \end{figure}

For this experiment, the cross-FROG traces of the first anti-Stokes Raman order were recorded with different instantaneous frequency separations by moving the mirror M7 backward and forward, which modifies the time delay between the pump pulses, which in turn modifies the instantaneous frequency separation. Fig.~\ref{fig: timedelay scan} shows the reconstructed pulses from the FROG algorithm. As can be seen from the spectra, strong red-shifting occurs when the instantaneous frequency separation is equal to or less than the Raman frequency 23.25 THz, as in Fig.~\ref{fig: timedelay scan}m and Fig.~\ref{fig: timedelay scan}n. As the instantaneous frequency separation decreases, the center frequencies of the two peaks get further apart until the resonance condition in Fig.~\ref{fig: timedelay scan}n. The instantaneous frequency dip is more pronounced for time delay cases -1/3~ps, 0~ps, and 1/3~ps.  For a given time delay, the different iterations in the experiment show that the instantaneous frequency is changing shape between iterations. This indicates a more dynamic double pulse system that's beyond the scope of this work. With more systematic and extensive data collection, it should be possible to explore these dynamics in greater detail and identify underlying trends.

\begin{table}
\centering
\caption{\centering Simulation results for 11 traces with the largest red-shifted spectrum}
\begin{tabular}{lllllllll}
\toprule
Trace & \thead{Error} & \thead{E1 \\ duration \\ (fs)} & \thead{E2 \\ duration \\ (fs)} & \thead{Amplitude \\ ratio} & \thead{Frequency \\ separation \\ (THz)} & \thead{Intensity\\ phase factor \\ (10$^{-3}$)} & \thead{Overall \\ time delay \\ (fs)} & \thead{Extra \\ phase} \\
\midrule
1 & 0.073 & 501 & 639 & 1.02 & 0.15 & 10.3 & 2749 & 3.23 \\
2 & 0.070& 534 & 783 & 1.12 & 0.44 & 3.98 & 2748 & 3.00 \\
3 & 0.065 & 809 & 1445 & 0.17 & 0.62 & 2.77 & 2585 & 2.32 \\
4 & 0.055 & 804 & 908 & 1.47 & 0.27 & 2.41 & 3035 & 1.59 \\
5 & 0.055 & 796 & 897 & 1.39 & 0.23 & 9.53 & 3036 & 3.01 \\
6 & 0.101 & 556 & 1090 & 0.85 & 0.67 & 0.02 & 2674 & 3.18 \\
7 & 0.099 & 532 & 1301 & 0.68 & 0.00 & 8.68 & 2997 & 2.77 \\
8 & 0.065 & 564 & 918 & 0.66 & 0.58 & 5.56 & 2661 & 3.21\\
9 & 0.096 & 591 & 700 & 1.04 & 0.36 & 10.79 & 2732 & 2.39 \\
10 & 0.094 & 768 & 875 & 1.28 & 0.30 & 2.37 & 2892 & 5.10 \\
11 & 0.068 & 552 & 801 & 0.76 & 0.57 & 7.28 & 2627 & 3.03 \\

\bottomrule
\end{tabular}

\label{tab:11traces}
\end{table}

\begin{figure}[ht]
    \centering
    \includegraphics[width=0.9\textwidth]{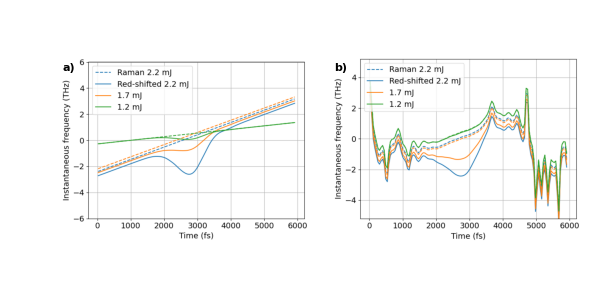}
    \caption{Simulation results from (a) beta-I model and (b) beta-I model with pump data. Instantaneous frequency traces for energy 1.2~mJ, 1.7~mJ, and 2.2~mJ are represented by different colors, with dotted lines indicating the Raman pulse and the solid lines indicating the Red-shifted pulse. }
    \label{fig: compare}
\end{figure}

Table~\ref{tab:11traces} shows the simulation results for 11 traces with the largest red-shifted spectrum. This lists all the parameters obtained from the simulations. As can be seen from the table, the average pulse duration for the Raman pulse $E_1$ is around $637$ fs, and the pulse duration for red-shifted $E_2$ is about $941$ fs. The differences between the experiments indicate that our laser system does vary from pulse to pulse.
The amplitude ratio between the red-shifted pulse and the Raman pulse is greater than 1, meaning that most of the energy is transferred to the red-shifted part. 
Fig.~\ref{fig: compare} shows the instantaneous frequencies of the anti-Stokes Raman spectra for the 3 traces discussed in Fig.~\ref{fig: energy scan frog} for both the beta-I model and the beta-I model with pump data. As shown by the dotted lines in the left plot, the instantaneous frequency of the Raman spectrum is a linear incline, while the red-shifted spectrum has a dip in the middle. This picture matches the idea of the intensity-dependent red-shifting. As observed at half of the pulse duration time, the higher intensity of the pumps will lead to a larger red shift. The trend is also reproduced for the simulation results shown in Fig.~\ref{fig: compare}b. Besides, the right plot shows that the linear chirp rate of the Raman part is similar to the chirp rate of the pump pulse. 

The formulation in Eqs.~(\ref{eq:omega1})-(\ref{eq: omega'}) indicates that variations in pump pulse energy modify the generalized Rabi frequency $\Omega'$ through simultaneous changes in the coupling amplitude $\Omega$ and the detuning $\Delta$. An increase in pulse intensity enhances the coupling strength and modifies the effective detuning, resulting in a larger value of $\Omega'$. In a two-photon dressed two-level system, temporal variations in the Rabi frequency predominantly influence the red-shifted spectral component, while the Raman spectrum itself remains largely unaffected. As a consequence, the instantaneous frequency of the Raman signal closely follows that of the pump pulses, whereas the red-shifted component exhibits a characteristic dip near the pulse center arising from the time-dependent modulation of the Rabi frequency.

\section{Conclusions}\label{sec5}
In this work, FROG is employed to characterize and interpret the red-shifted features appearing on the Raman orders. We systematically examine the evolution of the red-shifted spectral component as a function of both the pump pulse energy and the instantaneous frequency separation between the pumps. The FROG measurements reveal that pronounced red-shifted shoulders emerge when the instantaneous frequency spacing approaches the intrinsic Raman frequency of the medium. Under these conditions, the spectral structure of the first anti-Stokes order undergoes a clear transition, evolving from a cluster of closely spaced, unresolved features into a well-defined double-peaked profile.
In the pump pulse energy dependence study, higher pump energy leads to red-shifted shoulders with a larger frequency shift. To gain further insight into the underlying mechanism, each Raman order is modeled using a double-pulse representation. Simulations based on this double-pulse framework, incorporating both the Raman component and the accompanying red-shifted contribution, reproduce the experimentally observed features with a high level of consistency. 
In the instantaneous frequency view, the Raman pulse has a linear chirp with a similar chirp rate to the pumps, while there is a dip in the middle of the red-shifted pulse. We think the intensity-dependent Rabi frequency from two-photon dressed states would be the reason why we have the red-shifted Raman spectrum in the transient regime, MRG.

\section{Author Contributions}\label{sec6}
SS initiated the collaboration with DS and ZX.
DS and ZX led the experimental aspect of this research paper. SS and KKS trained bachelor students in this field at IIT Bhubaneswar and NIT Karnataka in India: students who contributed to the early development of this project. BDC significantly improved the existing early codes, contributed to code/algorithm development and model development, and the theoretical formulation for this research paper. PS worked as a guest researcher with SS and KKS at IIT Bhubaneswar. PS and SS contributed to writing the Introduction, Modeling Methodology. PS wrote the Appendix sections of the paper and developed the flowcharts to illustrate the implementation of the algorithm used in the simulations. SPA, SA, and SS contributed to the numerical implementation of various schemes, contributed to code development, code review, resolving bottlenecks, visualizing results, and writing this research paper. AP, KKS, and FB contributed to the internal review of the manuscript. SS managed the project from start to finish as the PI of this research project.

\section{Acknowledgements}\label{sec:ack}
SS and KKS trained bachelor students in this field at IIT Bhubaneswar and NIT Karnataka in India. During the early phases of this project, the curiosity-driven questions and spirited learning efforts by these bachelor students, Pravan Omprakash (NITK), Yash Vardhan (IIT BBS), and Shubhangi Sinha (IIT BBS), built the platform for senior researchers to develop upon. SS thanks Phani Motamarri and KV Vijay Girish at the Indian Institute of Science (IISc) Bangalore for valuable discussions on this project. SS and KKS acknowledge funding from MHRD and the Ministry of Steel in India. SS and FB gratefully acknowledge funding support from Deutsche Forschungsgemeinschaft (DFG) in Germany. SS and FB thank Prof. Bernd Markert, director at the Institute of General Mechanics at RWTH Aachen, for institutional support.

\section{Disclaimer}

BC worked on this project in a personal capacity. The views and conclusions expressed are solely those of the authors and do not reflect those of Meta. This work is independent of BC's professional duties and is neither affiliated with nor endorsed by Meta.

\section{Code/Data Availability}\label{sec8}

The code and the data sets for this research paper can be obtained by emailing the corresponding author, S. Swayamjyoti, at, swayamjyoti@iam.rwth-aachen.de OR swayam.sj@iitbbs.ac.in.

\section{Conflict of Interest}\label{sec9}

The authors declare no conflict of interest. 

\section{Appendix}\label{sec:appendix}

\subsection{Details of the developed framework}

In the entire framework of codes represented by the flowcharts, some key optimization terminologies are used. In order to achieve a better understanding of the codes, it is essential to understand these terminologies too (listed below). 
Original trace: Pixelized representation of the experimental trace (obtained from the FROG algorithm). 
\newline
\underline{Modules}: Files with the ". py" extension (containing Python code) that can be imported inside another Python Program. The modules are developed to perform a specific operation pertaining to the gradient descent optimization to compute the simulated trace. 
\newline
\underline{Diff\_array}: This variable contains the difference between the simulated trace and the experimental trace, which is further normalized with the value of the experimental trace. It's computed in the code diffarray.py.
\newline
\underline{Normalize}: The normalize.py makes the values of the simulation parameters positive if they are negative. Also, it adds a pi value to the phases.
\newline
\underline{Output}: A Python object computed by the entire code framework that contains information comprising A) scaled values of the training parameters B) Unscaled values of training parameters C) Diff\_array D) Loss E) frogscale.
\newline
\underline{Training parameters}: The parameters for the stochastic gradient optimization.
\newline
\underline{Scaled values}: The scaled values of the training parameters in the model are computed by the scaling function as shown in the flowchart for paramclass.py. See \ref{subsec3}. These scaled values are used in the ADAM optimization. Scaling helps in convergence in the ADAM optimization.  
\newline
\underline{Loss}: It is computed as the mean of the square of the difference between the simulated trace and the experimental trace. 
\newline
\underline{Frogscale}: It is the ratio of the maximum values in the tensor of simulated trace and experimental trace. 

\subsection{Framework of all codes \& All runs.py}

Our developed code, AllRuns.py initiates the simulations for the stochastic gradient descent optimization using the experimental trace files (containing the pixel values) as input. 
At the onset of this code, it imports all the relevant modules pertaining to the stochastic gradient descent optimization.
The AllRuns.py determines the number of input files from the provided folder paths, i.e., a total of 55 time delay scans in our case. Later, it runs (by calling the trainer.py module) the stochastic gradient descent optimization for each of these time delay scans one by one to generate the corresponding simulated trace. During this process, the iterative parameters like total number of iterations and post-processing details are also inputted to trainer.py. 
In the end, it obtains the output (details mentioned in common terminologies) in an object file for each of the time delay scans one-by-one in return from the trainer.py file.

\begin{figure}[ht]
    \centering
    \includegraphics[width=1.0\textwidth]{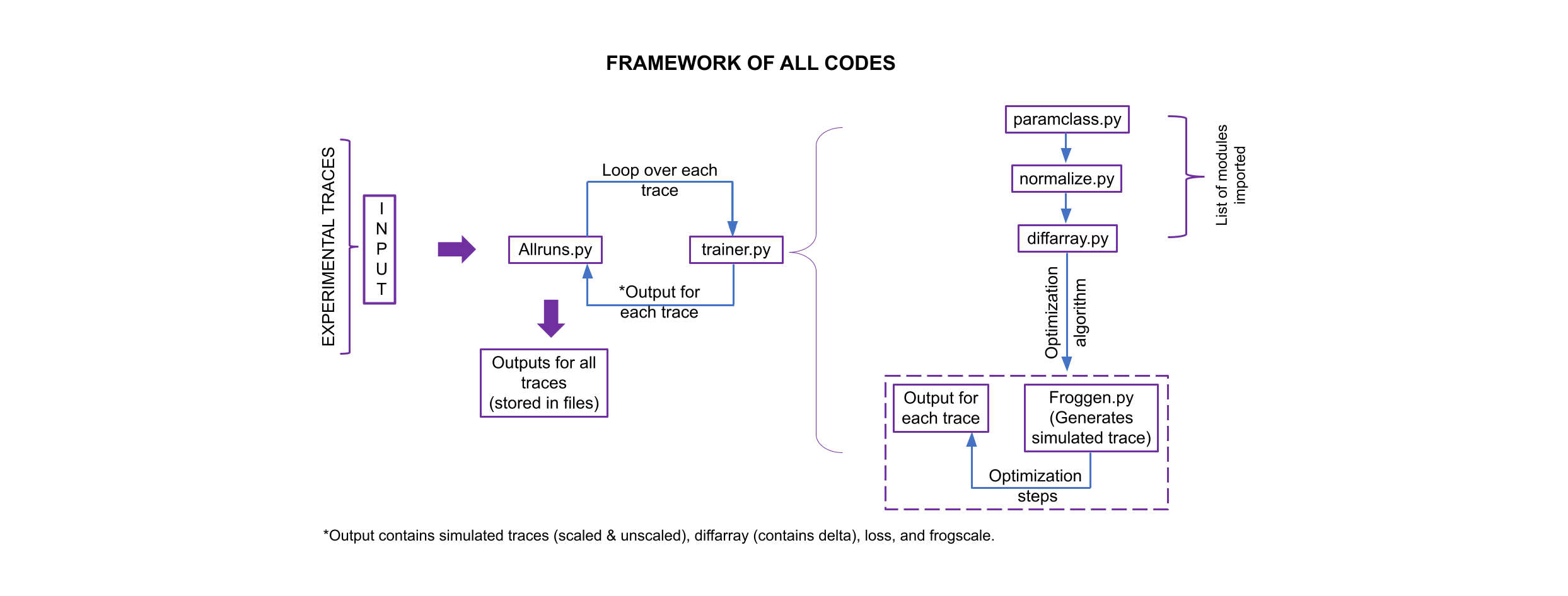}
    \caption{Overview of the Python codes.}
\end{figure}

\begin{figure}[ht]
    \centering
    \includegraphics[width=1.0\textwidth]{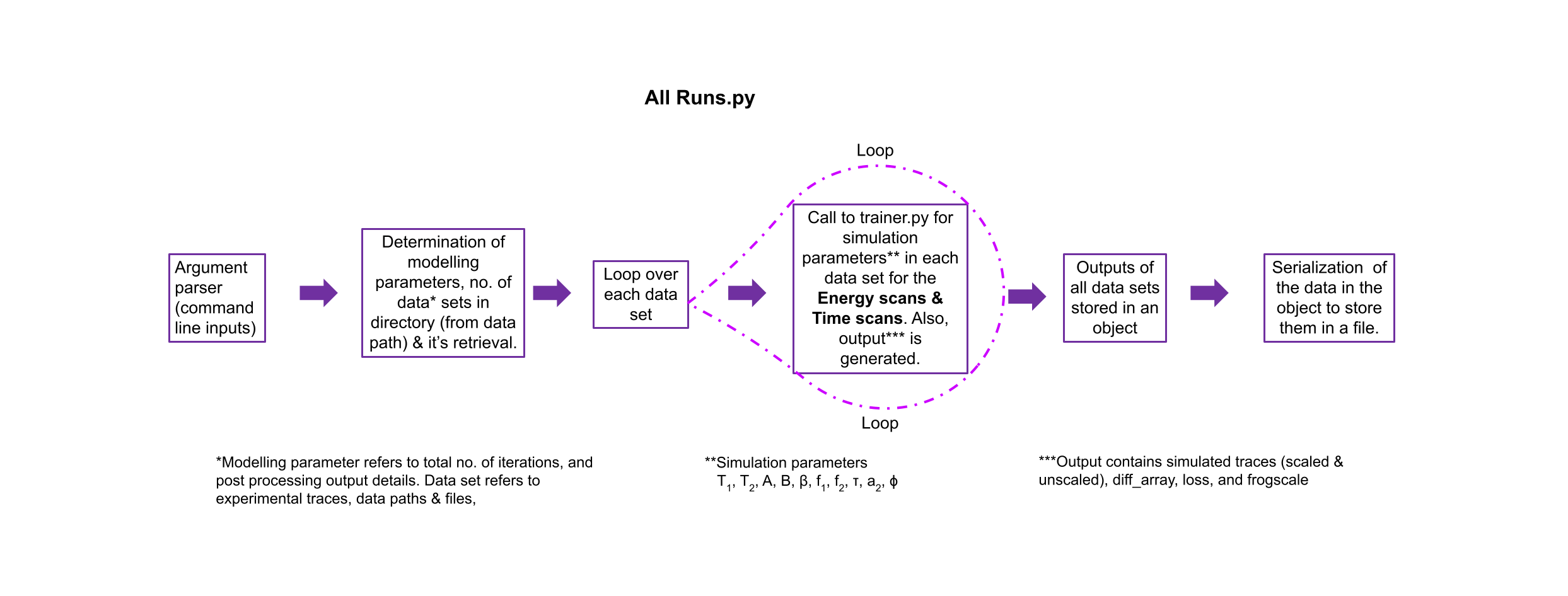}
    \caption{Structure of AllRuns.py}
\end{figure}

\subsection{Trainer.py}\label{subsec1}

This is one of the important modules in our framework of developed codes because it implements the gradient descent optimization algorithm. After the definition and initialization of all the parameters for optimization, it computes the normalization of some of the parameters and the norm of the experimental trace. This completes the precursor to the optimization. 
With the number of iterations defined, the optimization begins for the particular experimental trace file (one of the 65 experimental traces) received from trainer.py. 
\newline
At the first step of the optimization, the code invokes the froggen.py module to compute the simulated trace. Further, it executes the entire sequence of steps in optimization for each iteration, as shown in the flowchart, and generates the output. Later, it stores the output for all the iterations in an object. Finally, it provides this output object to the AllRuns.py for the inputted experimental trace file. 

\begin{figure}[ht]
    \centering
    \includegraphics[width=1.0\textwidth]{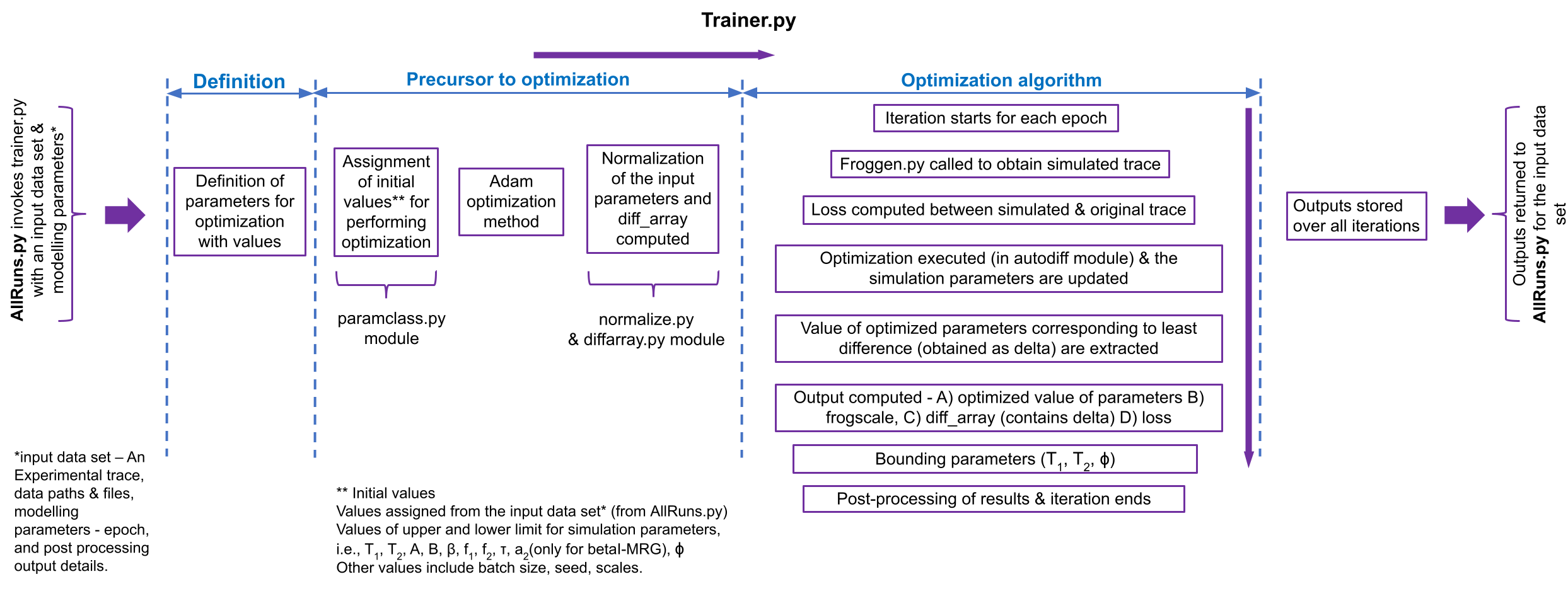}
    \caption{Structure of Trainer.py}
\end{figure}

\subsection{Froggen.py}\label{subsec2}

The optimization algorithm section of the trainer.py invokes froggen.py is invoked with input values consisting of the training parameters to construct the energy field for a time delay of $\tau$. 
\newline
Since the energy field is described by a Gaussian pulse function, these functions are also computed in the code. Next, the phases of the pump and Stokes pulses are computed. Subsequently, the energy fields for both these phases are computed. 
\newline
Finally, the Fast Fourier Transformation algorithm is used to convert the trial energy field into a simulated trace, as shown in the flowchart. Further, the simulated trace is passed to trainer.py to perform the optimization. 
 
\begin{figure}[ht]
    \centering
    \includegraphics[width=1.0\textwidth]{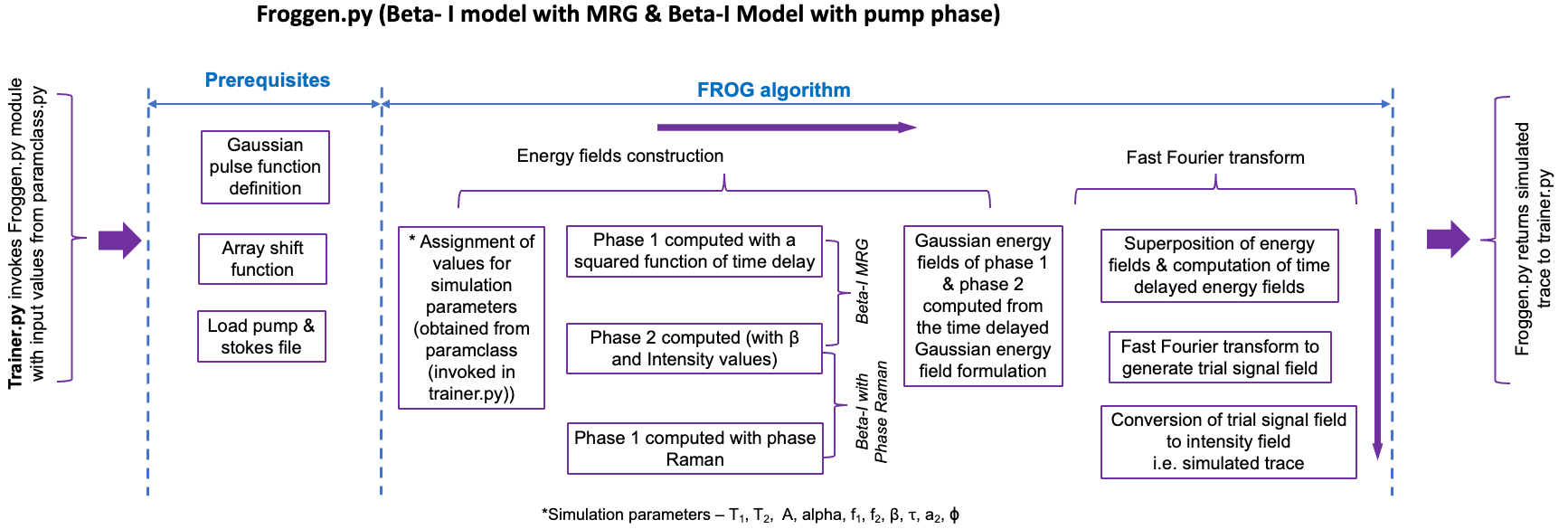}
    \caption{Structure of Froggen.py}
\end{figure}

\subsection{Paramclass.py}\label{subsec3}

The main objective of this module is to compute the scaled and unscaled initialized values of the parameters essential in the optimization procedure. Finally, it returns these values to the trainer.py.
\newline
Initially, trainer.py passes the range of values within which the values of these training parameters can be generated. 
\newline
Later, in order to obtain the scaled values of the training parameters, a scaling factor is adopted for each of them. For details, please refer to the flowchart. 
Finally, it returns these values to the trainer.py for performing optimization with the set of values for the parameters.

\begin{figure}[ht]
    \centering
    \includegraphics[width=1.0\textwidth]{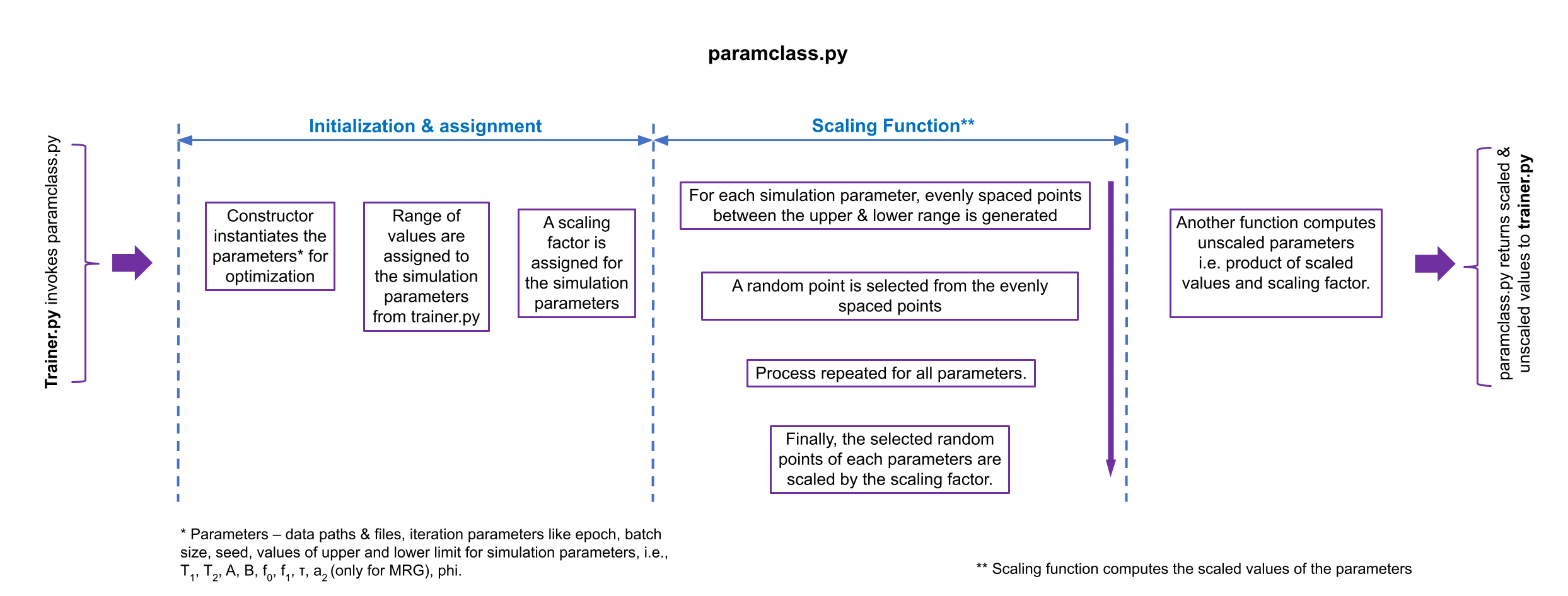}
    \caption{Structure of paramclass.py}
\end{figure}


\backmatter

\bigskip


\newpage

\bibliography{sn-bibliography, references}
\end{document}